\documentclass[numberedappendix,12pt,preprint]{emulateapj}

\newcommand{\simgt}{\,\rlap{\lower 3.5 pt \hbox{$\mathchar \sim$}} \raise
1pt \hbox {$>$}\,}
\newcommand{\simlt}{\,\rlap{\lower 3.5 pt \hbox{$\mathchar \sim$}} \raise
1pt \hbox {$<$}\,}

\def\psone{PS1{ }}
\def\panstarrs{Pan-STARRS~1{ }}

\shorttitle{QSO Variability Selection}
\shortauthors{Schmidt et al. (2010)}

 \usepackage{color}			      
 \definecolor{midgray}{gray}{0.4}		
 \usepackage[pdftex,colorlinks=true,citecolor=midgray,linkcolor=midgray]{hyperref}

\begin{document}


\title{Selecting Quasars by their Intrinsic Variability}


\author{Kasper B. Schmidt$^{1}$, 
Philip J. Marshall$^{2,3}$,
Hans-Walter Rix$^{1}$, \\
Sebastian Jester$^{1}$,
Joseph F. Hennawi$^{1}$, 
\& Gregory Dobler$^{4,5}$
}
\affil{$^{1}$ Max Planck Institut f\"ur Astronomie, K\"onigstuhl 17, D-69117 Heidelberg, Germany}
\affil{$^{2}$ Physics Department, University of California, Santa Barbara, CA 93106, USA}
\affil{$^{3}$ Kavli Institute for Particle Astrophysics and Cosmology, Stanford University, PO Box 20450, MS 29, Stanford, CA 94309, USA}
\affil{$^{4}$ Center for Astrophysics, Harvard University, Cambridge, MA 02138, USA}
\affil{$^{5}$ Kavli Institute for Theoretical Physics, University of California, Santa Barbara, CA 93106, USA}

\email{kschmidt@mpia.de}




\begin{abstract}

We present a new and simple technique for selecting extensive, complete and
pure quasar samples, based on their intrinsic variability. We parametrize the
single-band variability by a power-law model for the light-curve structure
function, with amplitude $A$ and power-law index $\gamma$. We show that
quasars can be efficiently separated from other non-variable and variable
sources by the location of the individual sources in the $A$-$\gamma$ plane.
We use  $\sim$60 epochs of imaging data, taken over $\sim $5 years, from the
SDSS stripe 82 (S82) survey, where extensive spectroscopy provides a reference
sample of quasars, to demonstrate the power of variability as a quasar
classifier in multi-epoch surveys.
For UV-excess selected objects, variability performs just as well as
the standard SDSS color selection, identifying quasars with a completeness of
90\% and a purity of 95\%. In the redshift range $2.5<z<3$, where color
selection is known to be problematic, variability can select
quasars with a completeness of 90\% and a purity of 96\%.
This is a  factor of 5-10 times more pure
than existing color-selection of quasars in this redshift range.
Selecting objects from a broad $griz$ color box \textit{without} $u$-band
information, variability selection in S82 can afford completeness and
purity of 92\%, despite a factor of 30 more contaminants than quasars
in the color-selected feeder sample. 
This confirms that the fraction of quasars hidden
in the ``stellar locus'' of color-space is small.
To test variability selection in the context of Pan-STARRS~1 (PS1) we
created mock PS1 data by down-sampling the S82 data to just 6 epochs
over 3 years. Even with this much sparser time sampling,
variability is an encouragingly efficient classifier. For instance,
a 92\% pure and 44\% complete quasar candidate sample is
attainable from the above $griz$-selected catalog.
Finally, we show that the presented $A$-$\gamma$ technique, besides
selecting clean and pure samples of quasars 
(which are stochastically varying objects), is also efficient
at selecting (periodic) variable objects such as RR Lyrae.

\end{abstract}



\section{Introduction}
\label{sec:intro}

Large, complete and pure samples of quasars have proven crucial for
observational cosmology. Quasars serve as probes of galaxy evolution, map
black hole growth and probe (and affect) the intergalactic medium. 
Quasar clustering is a tracer of mass clustering on both large
and small scales \citep{croom05,croom09b,shen07,shen09,ross09}, and the large samples provide
precise measurements of the evolution and spectral properties of the quasars
themselves \citep{vandenberk01,richards02b,richards04,richards06,richards09,boyle00,croom09b}. 
Furthermore, huge quasar samples are required to find a large number of gravitationally lensed
quasars \citep{oguri06,inada08}. Through the gravitationally magnified quasars
the quasar samples indirectly contribute to the understanding of the molecular
gas content in distant galaxies \citep{yun97,riechers07a,riechers07b}, mapping
of the intergalactic medium and structures 
\citep[e.g.][]{metcalf05,hennawi06,hennawi07} and exploration of the dark matter
(halo) content of galaxies \citep{dalal02,bradac02,dobler06,maccio08}. In short, large
well-defined quasar samples are a cornerstone of observational cosmology.

Photometric quasar samples have recently grown to nearly a million objects 
\citep[850,000 actual quasars;][]{richards09}.
Despite these impressive catalog sizes the number statistics still limit the achievable
science in various cases; 
especially those where particular and hence rare geometric constellations of quasars are
needed. For instance a 3$\sigma$ detection of a luminosity-dependent 
quasar bias above $z\gtrsim$1.9 when analyzing the angular clustering of quasars, needs
an estimated sample size of at least 1,200,000 actual quasars \citep{myers07a,myers07b,myers09}.
Searches for binary quasars \citep{hennawi06,hennawi09,myers08} - which provide interesting knowledge 
about small scale clustering and hence shed light on quasar triggering mechanisms 
and the nature of quasar progenitors - also should be based on samples with $>10^6$ actual quasars
in order to obtain reasonably sized statistical samples of possible quasar pairs. 
Also quasar-galaxy clustering \citep[e.g.][]{scranton05,lopez08,padmanabhan08,burbidge09}, 
i.e. exploring the statistics of quasars behind the foreground galaxies, calls for larger 
(relatively low-$z$) quasar samples than exist to date.
Furthermore, exploring the "transverse proximity effect" in the Ly$\alpha$ forest of quasars,
with foreground quasars near the sight line of background quasars \cite[e.g.][]{hennawi06a,hennawi07} 
is presently limited by quasar sample sizes. Obtaining larger photometric quasar catalogs 
to boost possible candidates for spectroscopic follow-up is needed.
The $\sim$3$\sigma$ detections of the integrated Sachs-Wolfe effect by cross correlating quasars
with the CMB to estimate the size of cosmological parameters and the dark energy
equation of state \citep[e.g.][]{giannantonio08,xia09,scranton05}
will also be improved by larger photometric samples of 1$<$$z$$<$5 quasars.
Last but not least, larger photometric quasar catalogs will enhance the number of known gravitationally 
lensed quasars \citep[e.g.][]{oguri10}. At present $\sim$100 quasar lenses are known and an even larger 
sample of the relatively rare gravitationally lensed quasar systems will among other things improve our 
knowledge about cosmology, galaxy mass distributions, quasar hosts and the growth
of the host's central black holes \citep[][and references therein]{schneider06}.
These few examples serve as a scientific justification for pursuing even 
larger photometric samples of low as well as of high redshift quasars. 

Most existing large quasar samples haven been selected on the basis of their
UV/optical colors or radio flux. However,
quasars are known to vary intrinsically on timescales of months to years and can
therefore be selected alternatively on the basis of their variability (alone).
Several physical processes are discussed as important causes of this
variability: foremost are accretion disc instabilities
\citep{rees84,kawaguchi98,pereyra06} but also large-scale changes in the
amount of in-falling material may be important
\citep[e.g.][and references therein]{hopkins06}.
Also, starbursts in the host galaxy \citep{aretxaga97,cidfernandes97}, micro
lensing by the host galaxy and compact dark matter objects
\citep{hawkins96,zackrisson03}, and stochasticity of multiple supernovae
\citep{terlevich92} have all been proposed. 
Irrespective of the physics behind the
variability, quasars are observed to exhibit
brightness variations, of typically  $\gtrsim10\%$ over
several years 
\citep[e.g.][]{giveon99,vandenberk04,rengstorf04,sesar07,macleod08,bramich08,wilhite08,kozlowski09,bauer09,kelly09}. This variability 
has been exploited for several purposes, e.g. to
estimate Eddington ratios and black hole masses \citep{bauer09,wilhite08}, or
simply to identify them \citep{geha03}. 

With SDSS, QUEST and OGLE \citep[see
e.g.][respectively]{abazajian09,rengstorf04,udalski97}, 
large-scale, multi-epoch and multi-band surveys
have emerged, and have been used to search for quasars. The Panoramic Survey
Telescope \& Rapid Response System 1 \citep[Pan-STARRS~1,][]{kaiser02} and 4
\citep[Pan-STARRS~4,][]{morgan08}, and the Large Synoptic Survey
Telescope \citep[LSST,][]{ivezic08,LSST} will take such surveys to the next level.
In all these surveys, 
the largest quasar samples stem from color selection in imaging
\citep[e.g.][]{richards02,richards04,richards09,atlee07,d'abrusco09}.  The characteristic 
so-called ``UV excess'' of quasars, their bright blue $u-g$ color, is
capable of separating the quasars from their stellar contaminants in 
color-color
space, allowing for efficient selection of targets for spectroscopic
follow-up \citep[e.g.][]{strauss02}.
Such UV excess color selection is, however, only efficient for low
($z\lesssim2.5$) and high ($z\gtrsim3$) redshift quasars, since the quasar and
stellar loci overlap in the $u-g$ color for $2.5<z<3.0$ objects, causing the
selection efficiency (or purity) in that region to drop below 50\%. 
For quasars with $2.6<z<2.8$ the
efficiency is close to 10\% \citep{richards06}. This confusion reigns until
the Ly-break of high-$z$ quasars moves into the $g$-band and again makes for
unusual colors \citep[e.g.][]{fan01,fan06}. 

Moreover, $u$-band imaging is expensive: the area and depth of an
optical imaging survey can be greatly increased by focusing on redder
filters, where atmospheric attenuation is lower and detectors more
efficient.  For example, the Pan-STARRS~1 telescope offers the
possibility of creating the largest sample of quasars to date with its
multi-epoch 30,000 deg$^2$, 3/4 sky, ``3$\pi$'' {\it grizY} imaging
survey.  For the purposes of identifying quasars in this data set, the
question remains, ``can we compensate for the lack of $u$-band data by
exploiting the multi-epoch nature of the $g$-band imaging instead?''
With one eye on the potential of Pan-STARRS~1, we therefore explore here the
possibilities of creating large, complete and pure samples of quasars
based on limited color information, but with light curves spanning
several years.  We use Stripe 82 of SDSS \citep{abazajian09} as a
testbed, both for the method in general and for making mock PS1 data
sets.

This paper is organized as follows. In Section~\ref{sec:var} we briefly review
previous attempts to characterize quasar variability in optical imaging
surveys, and then introduce the SDSS Stripe 82 data sets in
Section~\ref{sec:data}. We introduce our methodology for quantifying the
variability of various objects via their individual power-law structure
functions in Section~\ref{sec:pwrfit}, and show results of selection
experiments in Stripe 82 in Section~\ref{sec:results}. After a brief discussion in 
Section~\ref{sec:disc}, we conclude in Section~\ref{sec:conc}.  
All magnitudes are given in the AB system.


\section{Variability Characterization of Sources and the Structure Function}
\label{sec:var}

In this section we briefly review the strengths and weaknesses of optical
color selection, of particular sources, focusing on quasars.  We then present
our approach to quantifying source variability, which we will then explore as
an additional approach to selecting quasars, of other sources. 


\subsection{Color Selection}
\label{sec:COL}

\begin{figure}
\epsscale{1.15}
\plottwo{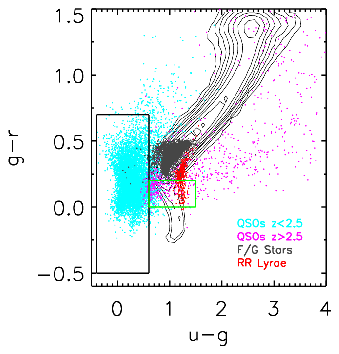}{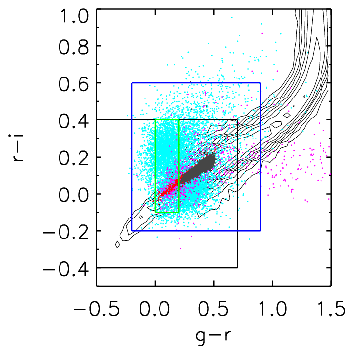}
\plottwo{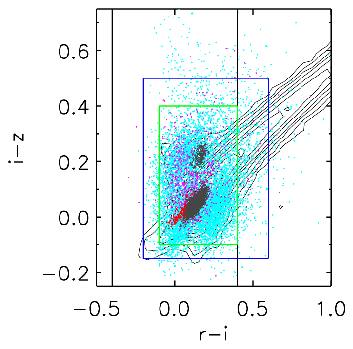}{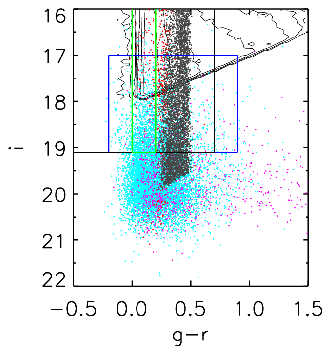}
\caption{Projections of point-source colors from the SDSS Stripe 82 data
described in Section~\ref{sec:data} to the $ugriz$ color space. The light blue
(magenta) points show $z<2.5$ ($z>2.5$)  spectroscopically confirmed quasars.
Illustrative 
contaminant point sources are shown as grey (F/G stars) and red (RR~Lyrae)
points. These panels demonstrate the importance of the $u$-band data in color
selection of quasars, with the $u-g$ color 
allowing the clearest discrimination. This clearly
shows the necessity for an alternative way to lower the amount of
contamination when the filters are too red. The lower right color magnitude
diagram illustrates that a cut in magnitude will also eliminate some
contaminants, and is included for comparison purposes with Figures~4, 7, 13 and
14 in \citet{richards02}. The black, green and blue boxes correspond to the
UVX, nUVX and $griz$ color boxes defined in Tables~\ref{tab:colorbox},
\ref{tab:colorboxNUV} and \ref{tab:grizbox} respectively. The narrow appearance
of the contaminant sample in the upper right panel is due
to our contaminants being mostly reference stars with small
photometric errors. The stellar locus spanned by all point sources in S82
with $15<r<18$ is shown as the black contours 
\citep[see also figures in][]{richards02}. }
\label{fig:coloroverlap}
\end{figure}

The most common way to generate large samples of optical 
quasar candidates for follow-up is
by specifying a particular region of interest in color space, as was
done e.g. in SDSS \citep{richards02,richards06,richards09}. For quasars at $z<3$ the $u-g$ color 
is crucial in this approach
since it enables a photometric separation of the quasar candidates from the stellar locus,
reducing the number of contaminating objects to a point where
spectroscopic follow-up is feasible. This is illustrated in Figure~\ref{fig:coloroverlap}, where we have
plotted the median color of $\sim$9000 spectroscopically confirmed quasars, as
well as an illustrative comparison sample of
5000 F/G and 483 RR~Lyrae stars (see Section~\ref{sec:data}), all drawn 
from the SDSS Stripe 82 photometric catalog DR7 \citep{abazajian09}. 
The top panels and the bottom left panel of Figure~\ref{fig:coloroverlap} shows the distribution of the
samples in the color-color planes of the SDSS $ugriz$ color cube. 
This clearly shows the power of the $u-g$ color (upper left panel) 
compared to the $g-r$, $r-i$ and $i-z$ colors 
in separating the quasars from their contaminants, especially 
for low redshift quasars (i.e. $z\lesssim2.5$ shown as
light blue points).
The color magnitude diagram in the lower right panel illustrates that a cut
in magnitude will also eliminate contaminants. The contours indicate the
stellar locus of Stripe 82 point sources with $15<r<18$

For higher redshift objects
($z\gtrsim 2.5$, shown as magenta points in Figure~\ref{fig:coloroverlap}) the
quasars intersect the stellar locus (top left panel). 
The purity for $2.5<z<3.0$ quasar candidate
samples is around 10-50\% in the color-selected SDSS quasar target sample
\citep{richards06}. 
In general the color
selection method is efficient for low and high redshift quasars,
but for the intermediate redshift objects contamination becomes a severe
problem.  We refer to Figures~13 and 14 in 
\citet{richards02} for a more complete
version of Figure~\ref{fig:coloroverlap}.

With \panstarrs (henceforth PS1) the contamination problem is even more pronounced
when using the color selection method only. \psone has a 5-filter system
consisting of SDSS-like $g$, $r$, $i$, $z$ bands (albeit with significantly
higher red sensitivity) and a $Y$~filter.  The crucial $u-g$ color used in the
SDSS color selection method is not available: the contamination of a
color-selected \psone quasar sample will be a problem for $z>2.5$ as well as
for $z<2.5$. 
It is therefore necessary to find a way of separating the
majority of quasars from the contaminating stellar locus in order to obtain a
pure \psone quasar sample. The intrinsic variability of the quasars (and
their contaminants) is a very promising tool for doing this. 


\subsection{Source variability: power-law structure functions}
\label{sec:SF}

The structure function characterizes the variability of 
quasars (and the other sources) by quantifying the variability amplitude
as a function of the time lag between compared observations 
\citep{cristiani96,giveon99,eyer02,vandenberk04,devries05,rengstorf06}. 
For any object the observables for estimating the structure function are
the $\frac{N(N-1)}{2}$ data pairs, assuming $N$ light
curve data points,
describing the variability as the magnitude difference between 
two epochs $i$ and $j$, corrected for measurement errors, i.e., 
\begin{equation}\label{eqn:datapairs}
V_{i,j}(\Delta t_{i,j}) =  \Delta m_{i,j} - \sqrt{\sigma_i^2+\sigma_j^2} \: .
\end{equation}
Here $\Delta m_{i,j}$ is the measured magnitude difference between observation
$i$ and $j$. The $\sigma_i$ and $\sigma_j$ are the photometric errors on the
measurements and $\Delta t_{i,j}$ is the time difference between the two
observations. 
The quantity $V$ is defined like this so that its average, over 
a large number of data pairs, is an {\it estimator for the intrinsic 
standard deviation of the source magnitude.}  

At this point we note that 
$\Delta t_{i,j}$ usually refers to the time lag in the quasar rest frame.
However, computing this requires 
{\it a priori} knowledge of the quasar redshift, and when selecting objects 
in imaging-only surveys, we do not know the object
redshift.  Therefore, we work with time lags in the {\it observed frame.}
This necessary convention differs from most of the quasar variability
literature; we will make some comparisons in Section~\ref{sec:pwrfit}.

In previous analyses, the average $V$
has been calculated in $n$~time lag bins using
data pairs from many quasars, thus ``stacking'' the variability signal to
allow the properties of the quasar population to be probed.
Then
\begin{equation}\label{eqn:SFbin}
V(\Delta t) = \left\langle 
  \sqrt\frac{\pi}{2} 
  |\Delta m_{i,j}|-\sqrt{\sigma_i^2+\sigma_j^2}  
  \right\rangle_{\Delta t} \, ,
\end{equation}
where the average, $\left\langle \; \right\rangle_{\Delta t}$, is taken over 
all epoch pairs $i,j$, whose lag falls in the bin $\Delta t$.
The same approach can be taken in estimating the structure function of classes
of objects (e.g. quasars at a given luminosity and redshift bin) if, say, only two epochs
are available per object, but large samples exist (e.g. \cite{richards06,richards09}); in that case the 
$\left\langle \; \right\rangle$ in Equation~\ref{eqn:SFbin} becomes and ensemble average.
\citet{vandenberk04} and others find that the ensemble average
quasar structure function appears to follow an increasing power law with time
lag.

On the other hand, for the case where the light curve sampling of 
each object is high, we can compute the average $V(\Delta t)$ for 
an {\it individual object} \citep{eyer02}.
Binning the $\frac{N(N-1)}{2}$ data pairs from an object's $N$-point
light curve gives an estimate of $V(\Delta t)$ defined at each bin center.
This approach is computationally efficient, and provides a free-form view of
the object's structure function.  
However, in the case of relatively sparse sampled data (6
epochs over 3 years in the \psone $3\pi$ survey, see Section~\ref{sec:DS}), 
binning the data pairs to obtain
the structure function from Equation~\ref{eqn:SFbin}
to estimate the variability may not be the optimal approach. 

In Equation~\ref{eqn:SFbin}, 
both the noise and the intrinsic 
photometric variability are assumed (implicitly) to have a
Gaussian distribution \citep{rengstorf06}.  We can then extend this simple model
of quasar variability to include a power law increase in variability with time
lag. Drawing on the results from \cite{richards06} we propose a power-law model 
for the structure function given by
\begin{equation}\label{eqn:powerlaw}
V_\textrm{mod}(\Delta t_{i,j} | A,\gamma) = 
  A \left( \frac{\Delta t_{i,j}}{1 \textrm{yr}}\right)^\gamma.
\end{equation}
We can then fit this model to a given set of data, $(\Delta m_{i,j},\Delta t_{i,j})$, as follows. 
We write the likelihood for the power law parameters $A$ and
$\gamma$ as
\begin{equation}\label{eqn:L}
\mathcal{L}(A,\gamma) = \prod_{i,j} L_{i,j} \; ,
\end{equation}
assuming a set of independent magnitude differences as our data. 
Here $L_{i,j}$ is the likelihood of observing one
particular magnitude difference
$\Delta m_{i,j}$ between two light curve points separated by $\Delta t_{i,j}$.
Following the ensemble analyses referred to above, we 
assume an underlying Gaussian distribution of $\Delta m$
values and Gaussian photometric errors:
\begin{equation}
L_{ij} = \frac{1}{\sqrt{2\pi \textrm{V}_{{\rm eff},ij}^2}} 
   \exp\left( -\frac{\Delta m_{ij}^2}{2\textrm{V}_{{\rm eff},ij}^2} \right),
\end{equation}
Here, the effective (observed) 
variability $\textrm{V}_\textrm{eff}$ is
\begin{equation}\label{eqn:Veff}
\textrm{V}_{{\rm eff},ij}^2 = 
   V_{\rm mod}(\Delta t_{ij}|A,\gamma)^2 + \left(\sigma_i^2+\sigma_j^2\right),
\end{equation}
i.e., we propagate the photometric errors $\sigma_i$ and $\sigma_j$ by adding
them in quadrature to the variability ``error'' $V_{\rm mod}(\Delta
t_{ij}|A,\gamma)$

This approach can yield posterior probability distributions on the two model 
parameters, $A$ and $\gamma$. 
The amplitude $A$ quantifies the root-mean-square magnitude difference on a 1
year timescale, while $\gamma$ is 
the logarithmic gradient of this mean change in magnitude. 
We assign uninformative priors for the parameters
(uniform in the logarithm of~$A$, and uniform in the arctangent of $\gamma$ --
since $\gamma$ represents the slope of a straight line), 
and then explore the posterior
probability distribution for these two power law parameters 
via a simple Markov chain Monte Carlo (MCMC) code 
\citep{metropolis53,hastings70} 
as described in
Appendix~\ref{sec:mcmc1}. 
All we are doing is replacing $n$ binned structure function parameters
(the values of $V$ in each of the $n$ bins) with two parameters that define a
power law structure function, and then inferring these parameters instead of
constructing estimators for them.
We will show in Section~\ref{sec:pwrfit} that using
a power law model for the variability actually provides a good fit.


\section{SDSS Stripe 82: a Testbed for Variability Studies}
\label{sec:data}

Anticipating the results of Section~\ref{sec:pwrfit}, we note that to detect
and quantify intrinsic quasar variability will likely require multi-epoch data
spanning several years. Before surveys with facilities such as Pan-STARRS and
LSST become available, SDSS Stripe 82 \citep{abazajian09} forms an excellent 
training set and methodological test bed \citep[e.g.][]{sullivan05,sesar07,bramich08, frieman08}.
In this section we will describe
the various Stripe 82 data sets that we have created in order 
to test and illustrate the prospects of our algorithm.

The SDSS Stripe 82 region (henceforth S82) covers approximately 320 deg$^2$,
from right ascension around 290$^\circ$ to 60$^\circ$ in a $2.5^\circ$ wide
band on the celestial equator. Over eight observing seasons it has been
repeatedly  observed in the fall months, resulting in many epochs (typically
$\sim$60) in each of the 5 SDSS bands.  As the \psone $3\pi$~survey will
contain fewer epochs, we can ``down-sample'' the S82 object light curves to
simulate observations taken with \psone (albeit ones at lower angular
resolution and depth).

Relative to PS1, S82 does have  the advantage of $u-g$ color coverage, and
extensive bright object  spectroscopy. One can therefore construct quite pure
samples of quasars, RR~Lyrae and so on, that may serve as ground truth for our
variability selection. 

In the following subsections we describe these various subsamples in some
detail, and provide a brief overview here. We have selected all the
spectroscopically confirmed quasars in S82 together with a representative  set
of (stellar locus) contaminants, which contains non-varying  (type F/G stars)
as well as varying (RR~Lyrae) point sources to illustrate our method and
algorithm prospects.  These objects' photometry data are plotted in $ugriz$
color space in Figure~\ref{fig:coloroverlap}. To investigate the selection of
quasars by their colors, we define three color selection boxes and explore the
objects returned by each. One of these mimics the more limited color selection
possible with PS1: quantifying how the variability information then improves
the \psone quasar selection is one of the main goals of this paper. 

In the following subsections, we describe two preparatory steps for
turning the $\sim$~60 epoch S82 data into a testbed for color plus variability
based quasar selection in SDSS (S82) and PS1: first, we describe the
definition of various sub-sets of candidate objects; then we describe some 
technical steps ``cleaning'' the light curves and down-sampling the S82 data
to mimic PS1. 


\subsection{Spectroscopically-confirmed quasars in Stripe 82}
\label{sec:qsos}

Key to designing a quasar variability selection algorithm is an understanding 
of   the variability properties of objects that are indeed spectroscopically
confirmed quasars. We have selected all of these (both point sources and
extended objects) published in the SDSS DR5 quasar catalog \citep{schneider07}
that fall within S82.  There are 9157 spectroscopically confirmed DR5 quasars
in S82, spanning a redshift range from 0.08 to 5.09. These quasars have
$15.4<i<22.0$ with a mean of 19.5.  See \cite{schneider07} for the
corresponding numbers for the whole DR5 quasar catalog.


To get the multi-epoch photometry (light curves) for the 9157 quasars we
performed an SQL neighbor search in the S82 DR7 database, 
choosing a search radius of
0.5'' to minimize the light curve contamination from 
misidentified (spatial) neighbors.
This search on average yielded 60 epochs per object, after 
selecting only entries with
good \verb+BRIGHT+, \verb+EDGE+, \verb+BLENDED+, \verb+NODEBLEND+,
\verb+SATUR+, \verb+PEAKCENTER+,  \verb+NOTCHECKED+, \verb+INTERP_CENTER+ and
\verb+DEBLEND_NOPEAK+ flags (of which the first 5 are referred to as fatal and
the rest as non-fatal flags by \citet{richards02} -- see their Table 2 or
\citet{stoughton02} Table~9 for a description of the flags). 
For consistency
we applied these same flag checks on {\it all} object samples we drew from the
S82 catalog. 
We describe these other samples below.


\subsection{Stellar locus ``contaminants'' in Stripe 82}
\label{sec:stelcont}

To get a sample of typical non-variable stellar contaminants we used the
SDSS standard star catalog of 1.01 million non-variable point-source objects in S82 published
in \cite{ivezic07}. From that we created a set of F/G-star colored objects, 
presumably non-varying, by
applying a color-magnitude cut on the standard star catalog so that $0.2 < g-r
< 0.48$ and $14.0 < g < 20.2$ for all the objects. This is a suitable cut for
F/G-stars according to the SEGUE team \citep{yanny09}
and makes them potential quasar sample  contaminants because of their $g-r$
color (see Figure~\ref{fig:coloroverlap}). We took a randomly selected
subsample of 5000 objects from this catalog and did a neighbor search in S82
to get multi-epoch observations of these contaminants. We again used a search
radius of 0.5'' and again made sure that none of the flags listed in
Section~\ref{sec:qsos} were set.

To be able to test whether our algorithm is able to separate quasars from
known variable contaminants, we used the largest available sample of securely
identified RR~Lyrae within S82 \citep{sesar09}, which consists of 483 RR
Lyrae. 

We will refer to the F/G stars and RR~Lyrae catalogs collectively as the
``contaminants'' in the remainder of the paper.


\subsection{UV-excess (UVX) objects}
\label{sec:UVXdata}

We would also like to test our ability to detect quasars in the
absence of spectroscopic data. To this end, 
we defined three photometrically-selected samples of S82
objects, whose variability properties we will explore.

The first of these is defined by a three-dimensional $ugri$ color box 
in which the SDSS quasar sample is complete for extinction-corrected $i$
magnitudes brighter than 19.1 \citep{richards02}. This color box is given in
Table~\ref{tab:colorbox},  and is shown in black lines in
Figure~\ref{fig:coloroverlap}.  
Note that this selection uses the SDSS
$u$-band data: we extracted all point sources within S82 that obeyed these
``UV excess'' (UVX) criteria. This returned a catalog of 2912 UVX point
sources. 

\begin{table}[!h]
\centering{
\caption[ ]{The UV excess (UVX) color box.}
\label{tab:colorbox}
\begin{tabular}[c]{|rcl|}
\hline
-0.4  $<$& $u-g$ & $<$ 0.6\\ 
-0.5  $<$& $g-r$ & $<$ 0.7 \\
-0.4  $<$& $r-i$ & $<$ 0.4 \\
 & $i$ & $<$ 19.1 \\
\hline
\end{tabular}}
\end{table}



\subsection{Non-UV excess (nUVX) objects}
\label{sec:nUVXdata}

As a compliment to the UVX object sample defined above, 
where the color selection is known to
efficiently return quasars at high completeness, 
a catalog of ``non-UVX'' (nUVX) objects was created from a region
of $ugriz$ 
color space where color selection of quasars is known to have problems. 
The color box from which we selected these nUVX point sources is given in
Table~\ref{tab:colorboxNUV}, and shown as a green box in
Figure~\ref{fig:coloroverlap} \citep[and also Figure~7 of][]{richards02}. 
In this
particular color box, the quasar locus, 
containing mostly intermediate redshift
($2.5<z<3$) quasars, crosses the stellar locus. The color selection therefore
has efficiency as low as 10\% in this region of color space
\citep{richards06}. In the nUVX color box we find 3258 objects in S82. 

\begin{table}[!h]
\centering{
\caption[ ]{The non-UV excess (nUVX) color box.}
\label{tab:colorboxNUV}
\begin{tabular}[c]{|rcl|}
\hline
0.6  $<$& $u-g$ & $<$ 1.5\\ 
0.0  $<$& $g-r$ & $<$ 0.2 \\
-0.1  $<$& $r-i$ & $<$ 0.4 \\
-0.1  $<$& $i-z$ & $<$ 0.4 \\
 & $i$ & $<$ 19.1 \\
\hline
\end{tabular}}
\end{table}



\subsection{Quasar Candidate Color Selection without $u$-band Data}
\label{sec:grizbox}

To simulate approximately the anticipated \psone $3\pi$ survey light curves, we
defined a third color box suitable for a first cut of the \psone catalog.  The
main purpose of this selection (where no $u$-band information is used) is to
excise the quasar locus as it threads through the 3-dimensional $griz$ color
space.  However, part of the stellar locus also lies in this box. We restrict
ourselves to right ascensions between~0 and 20~degrees (enclosing a sixth of
the S82 area, $\sim$50 deg$^2$) in order to return a manageable catalog of
12,714 objects. The $griz$ box is indicated by the blue solid lines in
Figure~\ref{fig:coloroverlap} and is defined in Table~\ref{tab:grizbox}. The
magnitude cut of~19.1 is chosen to allow straightforward  comparison with the
UVX and nUVX boxes.

\begin{table}[!h]
\centering{
\caption[ ]{The $griz$ color box.}
\label{tab:grizbox}
\begin{tabular}[c]{|rcl|}
\hline
-0.2  $<$& $g-r$ & $<$ 0.9 \\
-0.2  $<$& $r-i$ & $<$ 0.6 \\
-0.15  $<$& $i-z$ & $<$ 0.5 \\
17 $<$ & $i$ & $<$ 19.1 \\
\hline
\end{tabular}}
\end{table}



\subsection{Eliminating light curve ``outliers'' in Stripe 82}
\label{sec:LC}

\begin{figure}
\epsscale{0.99}
\plotone{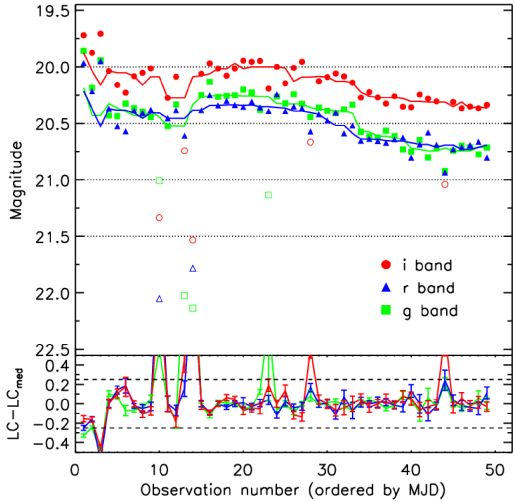}
\caption{Multi-epoch photometry output from SDSS Stripe 82 for the
spectroscopically confirmed quasar SDSS J203817.37+003029.8, 
shown in the  $g$ (green squares), $r$ (blue triangles) and $i$ (red circles)
bands. 
For a handful of epochs, the output magnitudes appear spuriously
faint; their inclusion would severely affect the calculation of a variability
structure function. Over-plotted are the corresponding medianized light
curves used to remove the outliers (open symbols) from the raw multi-epoch
output.  The dotted lines are plotted to guide the eye and are spaced by 
half a magnitude from 20 to 21.5.
In the lower panel the
residuals between the medianized light curve and the photometric measurements
for the three bands are shown, with the photometric errors over-plotted. The
limit used to remove the outliers ($|\textrm{LC}-\textrm{LC}_\textrm{med}| >
0.25$) is indicated by the horizontal dashed lines.}
\label{fig:LC}
\end{figure}

Plotting the complete S82 multi-epoch photometry output for the various
objects revealed some outlying points 
that were several magnitudes
fainter than adjacent flux points (see Figure~\ref{fig:LC}); 
only some of these outliers were found to be caused by image defects. However,
we assume that such a
significant decrease in magnitude in a single observation must be 
non-physical.
We therefore removed  the outliers (irrespective of their origin) by running a
median filter on the photometric measurements. Measurements with a residual
between the medianized light curve and the photometric data larger than
0.25~magnitudes  were removed. In Figure~\ref{fig:LC} we show the $g$, $r$ and
$i$-band multi-epoch photometric measurements (open symbols indicating the
removed measurements) with the corresponding medianized light curves
over-plotted for quasar SDSS J203817.37+003029.8. The bottom panel shows the residuals,
with the limit of 0.25~magnitudes indicated by the dashed lines. 
As is the case here, in general, only a small fraction of the epochs is
removed. 

It is these cleaned 
multi-epoch measurements, where the outlying observations have been
removed (i.e. the filled symbols in Figure~\ref{fig:LC}), 
we use in the determination and
exploration of the objects' variability. 


\subsection{Down-sampling S82 light curves to the \psone cadence}
\label{sec:DS}

In order to explore quasar selection in the context of the \psone $3\pi$ survey, we
down-sampled the S82 data to mimic the planned \psone observations 
\citep[][shown schematically in Table~\ref{tab:Pan-STARRS1obs}]{PS1}. 
We assumed 3 observing seasons for PS1, with a duration of 155
days (covering all filters) each. 
Only S82 objects with more than 7~epochs in each (SDSS) season were passed
to the actual down-sampling routine: $\sim$1\% of the quasars, $<0.1\%$ of the
F/G stars and $\sim$20\% of the RR~Lyrae did not satisfy this criteria. 

We down-sampled the S82 light curve data by matching each season of observations
for the $g$, $r$, $i$ and $z$-band with the (approximately) correct time
intervals between consecutive observations in each band. 
No color information went into the down-sampling.
After identifying 6~suitable S82 epochs in each
band we removed all other observations, providing a set of mock \psone
data. 

\begin{table*}[thbp]
\begin{center}
\caption{A schematic overview of the planned \psone $3\pi$
survey schedule.}
\label{tab:Pan-STARRS1obs}
\begin{tabular}[c]{|l|ccccccccccccccccccccccccccccccc|}
\hline
Date (days) & -60 & . & . & . & . & . & -30 & . & . & . & . & . & 0 & 5 & 10 & . & . & . & 30 & 35 & 40 & . & . & . & 60 & . & . & . & . & . & 90  \\
Band  & $z$ & . & . & . & . & . & $Y$ & . & . & . & . & . & $i$ & $r$ & $g$ & . & . & . & $i$ & $r$ & $g$ & . & . & . & $Y$ & . & . & . & . & . & $z$  \\
Moon & F & . & . & N & . & . & F & . & . & N & . & . & F & . & . & N & . & . & F & . & . & N & . & . & F & . & . & N & . & . & F  \\
\hline
\end{tabular}
\end{center}
\raggedright\footnotesize Notes: Date is calculated with respect to the first
$i$-band measurements. Band shows the band observed. Moon indicates
whether the moon is full (F) or new (N). The columns are spaced by five
days.
\end{table*}



\section{Power-Law Structure Functions for Sources in Stripe 82}
\label{sec:pwrfit}

\begin{figure}
\epsscale{0.99}
\plotone{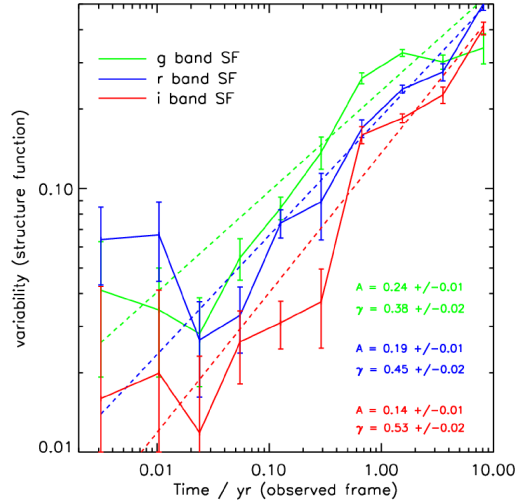}
\caption{Variability structure functions (solid lines) for SDSS J203817.37+003029.8, based on the photometry shown in Figure~\ref{fig:LC} for the $g$ (green), $r$ (blue) and $i$ (red) band, computed from Equation~\ref{eqn:SFbin}. The best-fit power law (Equation~\ref{eqn:powerlaw}) model for the structure function from the MCMC simulated annealing code (Appendix~\ref{sec:mcmc1}) are over-plotted as dashed lines. The corresponding $A$ and $\gamma$ parameters of the power law and their estimated errors are quoted in the lower right corner of the plot. Calculating similar structure functions by means of the power-law model for the 9157 Stripe 82 quasars give the median quasar sample structure function shown in the top panel of Figure~\ref{fig:meanStrAll}.}
\label{fig:SF}
\end{figure}

In Figure~\ref{fig:SF} we show the binned
structure functions (Equation~\ref{eqn:SFbin}) from the $g$, $r$ and $i$-band 
light curves of quasar SDSS~J203817.37+003029.8 (Figure~\ref{fig:LC}). 
Calculating binned structure functions for individual
S82 objects is a simple way of quantifying the variability of each object
if there is a large number of epochs available.
However, Figure~\ref{fig:SF}
suggests that we might indeed 
be justified in further compressing the
structure function into a two parameters power-law fit, 
as proposed in Section~\ref{sec:SF}.

\begin{figure}
\epsscale{1.}
\plotone{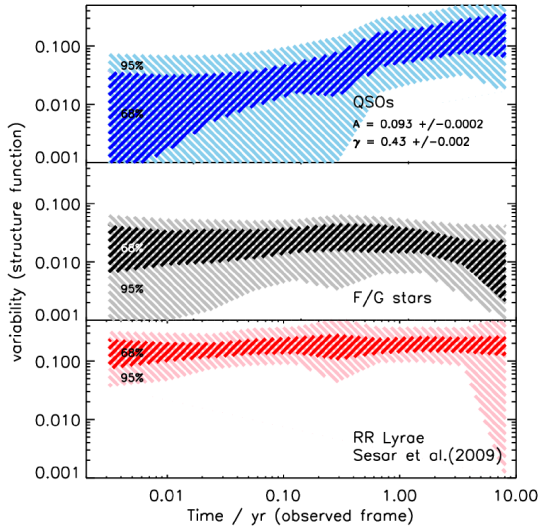}
\caption{The sample-median $r$-band binned
structure function for the quasars (top),
F/G stars (center) and 483 RR~Lyrae (bottom) in Stripe 82, 
calculated using Equation~\ref{eqn:SFbin}. The power-law nature
of the three samples is clearly seen,
by the approximate straight lines the
structure functions trace. 
The power law parameters $A$ and $\gamma$ obtained by 
fitting the quasar sample
structure function with the 2 parameter power law model
\ref{eqn:powerlaw} 
are shown in the top panel. The shaded regions
around each structure function indicate the 68\% and 95\% scatter around the
median value.}
\label{fig:meanStrAll}
\end{figure}

Before doing so, we explore wether this power law behavior is present in an
average  sense. In Figure~\ref{fig:meanStrAll} we show the median sample
structure function, created by median-combining all the individual binned
structure functions calculated with Equation~\ref{eqn:SFbin} separately for
the well sampled S82 quasars (Section~\ref{sec:qsos}),  F/G-stars
(Section~\ref{sec:stelcont})  and RR~Lyrae \citep[Section~\ref{sec:stelcont}
and][]{sesar09}. The shaded regions in Figure~\ref{fig:meanStrAll} around the
medians enclose 68\% and 95\% of the individual structure functions.
Figure~\ref{fig:meanStrAll} shows that the quasar sample median of the binned
structure function closely resembles a power law with $A=0.093\pm0.0002$ and
$\gamma=0.43\pm0.002$, in agreement with findings elsewhere in the literature
\citep{vandenberk04,rengstorf06,wilhite08,bauer09}.   In particular, a value
of the slope of the sample median structure function  of $\gamma = 0.43$
agrees well with most of the literature  (see e.g. Table~4 in \cite{bauer09}
for a brief overview).  Rough estimates of the 1-year observed frame power law
amplitudes in the literature give amplitudes between 0.10 and 0.14 (depending
on the assumed mean redshift of the samples), in good agreement with our
estimate for $A$ of 0.093~mag.

Figure~\ref{fig:meanStrAll} also shows that the sample structure  functions of
contaminants are also well described by power laws, but with small values of
$\gamma$  (i.e. they show no long-term growth in their variability).  Note
that the F/G-stars, chosen to be non-variable, have a variability amplitude
of~$\sim$0.04 mag. The RR Lyrae variability, when sparsely and randomly
sampled in S82,  looks like white noise ($|\gamma| \ll 1$) with an amplitude
$\sim$0.2 mag. Thus, the use of a  power-law model of the form given in
Equation~\ref{eqn:powerlaw} would seem to be  a fairly good assumption, and
different types of objects may differ both in amplitude and in slope of their
structure function.

In Figure~\ref{fig:SF} the power law fits to the three (binned)  structure
functions are shown as dashed lines. The $A$ and $\gamma$ with their estimated
errors are given in the lower right corner of the plot.

\section{Results}
\label{sec:results}

Having defined different sub-samples of sources, and having shown that we can
sensibly quantify their light curve characteristics by a power-law structure
function model, we now proceed to characterize each of these sources by
their best-fit parameters $A$ and $\gamma$.

As opposed to the earlier SDSS analysis \citep{richards02,richards06,richards09}
the improved time sampling of the S82 (and even PS1)
surveys, enables us to investigate the {\it distributions} of the $A$ and
$\gamma$ parameters for the individual sources, not only for ensembles.


\subsection{The A-$\gamma$ Distribution}
\label{sec:AGspace}

\begin{figure}
\epsscale{0.99}
 \plotone{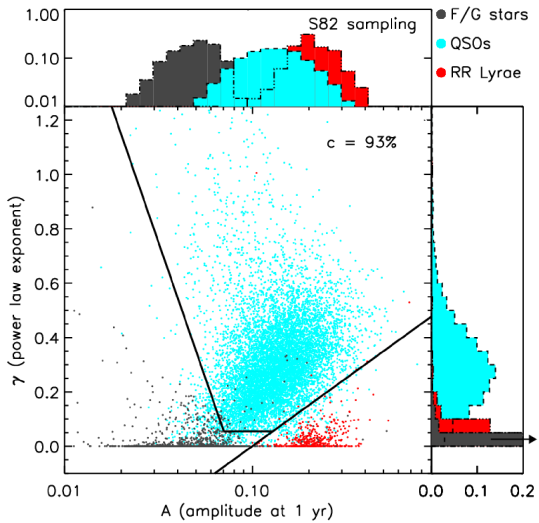}
\caption{Distribution of the variability structure function parameters $A$ and
$\gamma$ (Equation~\ref{eqn:powerlaw}) for $\sim$15,000 individual objects in Stripe 82.
The spectroscopically confirmed quasars are shown as light blue points;
confirmed RR~Lyrae and color-selected F/G stars are shown in red and grey
respectively. A separation of the quasars from the stellar locus contaminants
is clearly seen. The three solid lines
(Equations~\ref{eqn:cutDSone}--\ref{eqn:cutDSthree}) 
define the region in which we
estimate the quasar completeness~$c$ of our algorithm (which turns out to be
93\% in this case, Table~\ref{tab:compur}). Along the axes we show the
projected $A$ and $\gamma$ distributions for the sub-samples.}
\label{fig:AvsGallNOBIN}
\end{figure}

Figure~\ref{fig:AvsGallNOBIN} shows the distribution of the variability
characteristics, quantified by the best-fit $A$ and $\gamma$ for all the
spectroscopically confirmed quasars, and for the ``contaminant'' F/G stars
and  RR~Lyrae, as described in Section~\ref{sec:data}, based on their $r$-band
S82  light curves.  The histograms along the axes of the two dimensional
scatter plot show the projected parameter distribution of the quasars and of
the contaminants. Visual inspection of Figure~\ref{fig:AvsGallNOBIN} alone
shows how well the spectroscopically confirmed quasars separate from the
(stellar locus) contaminants in this space, demonstrating that the power-law
structure function fit from a single band is an efficient classifier for data
of this quality ($\sim$60 epochs). 

\begin{figure}
\epsscale{0.99}
 \plotone{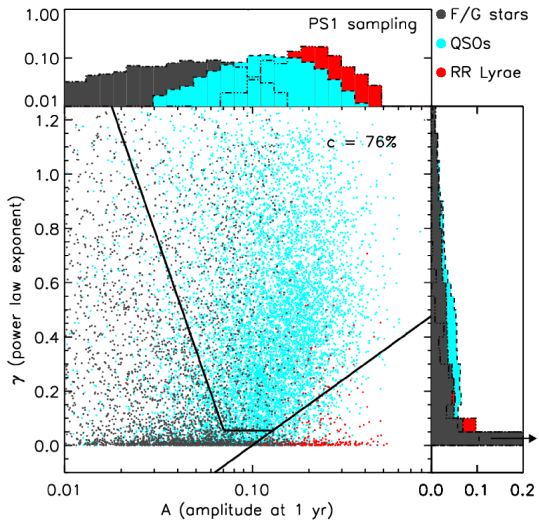}
\caption{Distribution of the variability structure function parameters, similar to
Figure~\ref{fig:AvsGallNOBIN}, but after down-sampling to the 6 epochs, expected
 \psone $3\pi$ survey cadence (Section~\ref{sec:DS}). 
A variability separation of quasars and their contaminants is again apparent,
albeit not as clearly as in Figure~\ref{fig:AvsGallNOBIN}. 
The selection region is the same as in Figure~\ref{fig:AvsGallNOBIN}; the
completeness  is given in the upper right corner of the plot, and in
Table~\ref{tab:compur}.}
\label{fig:AvsGdsNOBIN}
\end{figure}

The analogous $A$-$\gamma$ distributions for the much sparser PS1-like
sampling of the $r$-band measurements (Section~\ref{sec:DS}) are shown in
Figure~\ref{fig:AvsGdsNOBIN}: these parameter estimates are based on only 6
epochs of photometry over 3 years, rather than the $\sim 60$ in the full S82
survey.   The separation of the quasars and the contaminants is less
clean with the \psone sampling, but one nevertheless clearly sees a
quasar-dominated region with rather low contamination.
Plots analogous to the ones shown in Figures~\ref{fig:AvsGallNOBIN}
and~\ref{fig:AvsGdsNOBIN}, but  for the $g$, $i$ and $z$-band measurements,
show that on average $A$~decreases by 30-50\% going from $g$ to $z$ band,
whereas $\gamma$ is unchanged with varying wavelength.  Thus, the separation
of the quasars from the contaminants via their $\gamma$ values appears to work
comparably well in all 4 bands (with somewhat more scatter in the $z$-band).
In agreement with \cite{kozlowski09}, no clear difference in the ratio of the
amplitudes at different wavelengths between RR~Lyrae and quasars is detected.

Since most F/G-stars should not vary, but RR~Lyrae do, they should have
different $A$~distributions. This is seen in Figure~\ref{fig:AvsGallNOBIN} and
\ref{fig:AvsGdsNOBIN}: RR~Lyrae have magnitude amplitudes above $\sim$0.1
\citep[e.g.][]{soszynski03,sesar09}, while the F/G-stars have characteristic 
values of $A\sim0.01$. It is therefore clear that our approach can also
separate RR~Lyrae from stellar (non-varying) contaminants without doing a full
fit of a periodic light curve.


\subsection{Completeness and Purity}
\label{sec:compur}

To move beyond a merely qualitative assessment of the separability of
the quasars from contaminants we now estimate the achievable
completeness and purity of the resulting sample.  Purity is the more
difficult quantity to estimate as it requires appropriate abundances
for the contaminants. We do \textit{not} have these for the training
set parent quasar, F/G star and RR~Lyrae samples,  where the   ratio of
quasars to contaminants is 2:1, instead of a more realistic $\sim$1:25,
and therefore can only estimate completeness when working with these
samples.   However, by design the UVX, nUVX and $griz$ selection boxes
(Sections~\ref{sec:UVXdata}--\ref{sec:grizbox} and
\ref{sec:UVX}--\ref{sec:grizboxres})  give parent samples with the
correct quasar-contaminants ratio: we use these to explore the purity
of our quasar selection algorithm.

For the present we divide the $A$-$\gamma$ plane by simple cuts that
define a quasar selection box, and then quantify its performance. 
Specifically, our fiducial quasar selection region is bounded by  the
following three straight lines:
\begin{eqnarray}
\gamma(A) &=& 0.5*\log(A)+0.50  \label{eqn:cutDSone} \\
\gamma(A) &=& -2*\log( A)-2.25 \\
\gamma(A) &=& 0.055 \label{eqn:cutDSthree}
\end{eqnarray}
These cuts are shown as black solid lines in
Figures~\ref{fig:AvsGallNOBIN}--\ref{fig:AGcolorPS1}.

This can be thought of as a way of providing lower limits on the
available completeness and purity, that a more sophisticated selection
procedure would improve upon. One could of course tweak the cuts in 
Equations~\ref{eqn:cutDSone}--\ref{eqn:cutDSthree}  to explore the
trade-off between purity and completeness, which we have only done here
``by eye.''

Applying these cuts to the data leads to the completeness given in
Table~\ref{tab:compur} for the ``QSO+contam'' catalog. The completeness
is calculated as the fraction of spectroscopically confirmed quasars in
the sample that fall within the cuts.  In our simple illustrative setup
we have a completeness of 93\% for the quasars in the case where the
time sampling is equal to Stripe 82 ($\sim$60 epochs). In the case of a
PS1-like time sampling the completeness drops to 76\%. 

\begin{table*}[thbp]
\centering{
\caption[ ]{The completeness~$c$ and purity~$p$ of the SDSS Stripe 82
(S82) and mock \psone variability-selected object catalogs. The $\delta c$ and $\delta p$ indicate the poisson error on $c$ and $p$.}
\label{tab:compur}
\begin{tabular}[c]{|l|l|c|c|c|c|l|}
\hline
Parent Catalog  & Sampling  & $c$   & $\delta c$  & p     & $\delta p$ & Quasar Reference Catalog  \\
\hline\hline 
QSO+contam.     & S82       & 93\%  &  1\%        & -     & -          & SDSS                      \\
QSO+contam.     & PS1       & 76\%  &  1\%        & -     & -          & SDSS                      \\
\hline 
UVX             & S82       & 90\%  &  3\%        & 95\%  &  3\%       & SDSS+2SLAQ                \\
UVX             & PS1       & 73\%  &  2\%        & 92\%  &  3\%       & SDSS+2SLAQ                \\
\hline 
nUVX            & S82       & 90\%  & 15\%        & 96\%  & 16\%       & Visual                    \\
nUVX            & PS1       & 65\%  & 12\%        & 32\%  &  5\%       & Visual                    \\
\hline 
$griz$ box      & S82       & 92\%  &  6\%        & 92\%  &  6\%       & Visual \& SDSS+2SLAQ      \\
$griz$ box      & PS1       & 75\%  &  6\%        & 30\%  &  2\%       & Visual \& SDSS+2SLAQ      \\
\hline 
\end{tabular}}
\end{table*}


Of the RR~Lyrae 97\% and 83\%  (for the S82 and \psone sampling
respectively) fall in the high-$A$ low-$\gamma$ corner below the line
given by Equation~\ref{eqn:cutDSone}. In this region $<$0.5\% of the
5000 F/G-stars lie, illustrating quite a clean separation between
RR~Lyrae and non-varying stellar contaminants.

Besides calculating the overall completeness, we also split our data
into redshift bins. The completeness of the quasars is rather constant
as a function of redshift, with a minor loss of about 5-10\% for
redshifts above 4.  Since our training set only contains 52 quasars at
$z>4$, we were not able to investigate this decreased completeness
further.

In the next three subsections we proceed to explore the $A$-$\gamma$
distributions for the samples of objects that were selected only on the
basis of their colors (Table~\ref{tab:colorbox}, \ref{tab:colorboxNUV} 
and~\ref{tab:grizbox}). For those samples, we estimate the
completeness  (as above) and also the purity, defined as the fraction
of known quasars  compared to the total number of objects inside the
$A$-$\gamma$ selection regions.


\subsubsection{Quasars in the UVX catalog}
\label{sec:UVX}

For the UVX object sample (Section~\ref{sec:UVXdata})  there is enough
spectroscopy in S82 to define a spectroscopically confirmed quasar
subsample, a reference catalog of quasars which is complete in S82.  
By combining the spectroscopically confirmed quasars in S82
(Section~\ref{sec:qsos}), with the objects from the 2SLAQ (2-degree
field SDSS luminous red galaxies and QSO) survey\citep{croom09},  our
final quasar reference catalog contains 11216 individual quasars (9157
from SDSS and 2059 from 2SLAQ). We only selected objects flagged as
``QSO'' in the publicly available 2SLAQ
data.\footnote{http://www.2slaq.info/} 

Matching this quasar reference catalog with the catalog of UVX S82
point sources returned 2140 quasars out of the 2912 objects in the UVX
catalog. Thus, 73\% percent of the point sources in the UVX color box
are known quasars. It is not surprising that the fraction is so large,
since we used the powerful $u-g$ color in the definition of the color
box. This simply re-affirms that the UVX box is a region of color space
where the quasars are in the majority,  as they are well separated from
the stellar locus by the $u-g$ color; it is exactly this separation to
which we are trying to find an alternative.

\begin{figure*}
\epsscale{0.95} 
\plottwo{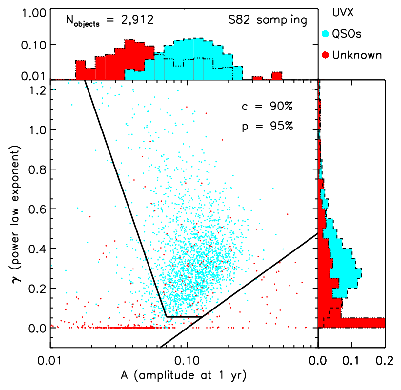}{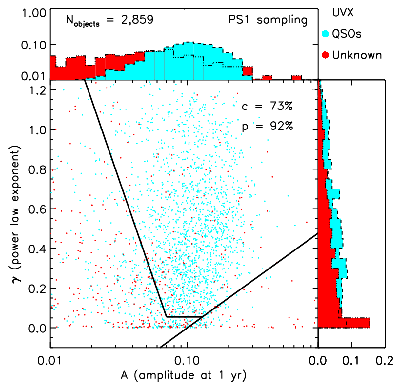}
\plottwo{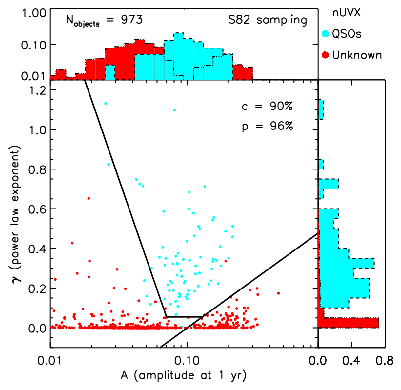}{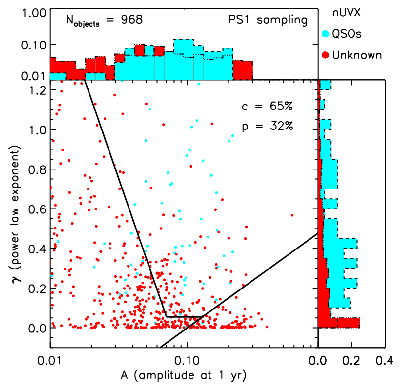}
\caption{Distribution of variability structure function power law
parameters $A$ and $\gamma$ measured for the objects in the UVX and
nUVX catalogs of Table~\ref{tab:compur}. The left plots corresponds to
the catalogs with a Stripe 82 time sampling, while the right plots
correspond  to the sparser \psone time sampling (6 epochs over 3
years).   The three solid lines in each scatter plot correspond to the
quasar variability selection region defined by
Equations~\ref{eqn:cutDSone}--\ref{eqn:cutDSthree}. The completeness
and purity  estimates are shown in the upper right corner of each
scatter plot, and in Table~\ref{tab:compur}.  In the upper left corner
the total number of objects in the catalog is stated. The light blue
points indicate the known quasars, and the red points all other
objects.  Along the axes of the two dimensional scatter plots the
projected probability distributions for $A$ and $\gamma$ are shown as
histograms.  }
\label{fig:AGs}
\end{figure*}

We estimated $A$ and $\gamma$ for the entire UVX catalog, using both
the full (S82) sampling and the sparser PS1-like version of it. The
result is shown in the top row of the $A$-$\gamma$ plots in
Figure~\ref{fig:AGs}. Applying our simple variability selection cuts
(Equations~\ref{eqn:cutDSone}--\ref{eqn:cutDSthree}) returned 2033 and
1734 quasar candidates for the S82 and PS1 sampling, respectively.
Matching these objects with the 2140 know SDSS+2SLAQ quasars in the UVX
catalog returned 1935 and 1573 matches. Thus we are able to detect the
UVX quasars with a completeness of 90\% and a purity of 95\% (1935
matches/2033 candidates) when using the S82 time sampled data. For the
PS1-like sampling of the data we  get a completeness of 73\% and a
purity of 92\% (Table~\ref{tab:compur}).

\subsubsection{Quasars in the nUVX catalog}
\label{sec:greenbox}

The catalog of the UVX point sources was deliberately chosen from a
region of color space where the color selection already does a superb
job finding quasars, as confirmed by the complete SDSS and 2SLAQ quasar
catalog and the completeness and purity of our $A$-$\gamma$ approach. 
However, one might argue that in this case (of UVX quasars) a light
curve analysis adds little. To explore the $A$-$\gamma$ approach
further we applied it to the non-UV excess objects described in
Section~\ref{sec:nUVXdata}.

In the nUVX color box (Table~\ref{tab:colorboxNUV}) there is no simple
way to quantify the completeness of the parent sample of color-selected
objects, since we do not know how many quasars were missed in this
region of color space during the SDSS survey. To try to quantify this,
we extracted the spectra from the 973 objects in  this catalog that
were targeted  for SDSS spectroscopy.  By visually inspecting these
spectra we were able to compile a catalog of 77 quasars among the S82
nUVX objects. (Of these 77, 6 quasars are not in the SDSS+2SLAQ
catalog).  
This means that if the SDSS fibers had been allocated to nUVX objects
randomly, then the purity of the nUVX quasar sample would be $77/973 =
8$\%. However, in practice the fibers were placed according to a
Bayesian ranking that made fuller  use of the color information, so
that this 8\% is likely an upper limit on the purity of the nUVX quasar
sample. (The fraction of quasars hidden in the un-targeted nUVX objects
is likely lower than the fraction of quasars found in the nUVX objects
with spectra.)

Applying our  $A$-$\gamma$ analysis and our variability selection
criteria to the spectroscopic sub-catalog of 973 nUVX objects returned
72 (178) quasar candidates for the S82 (PS1-like) time sampling. The
nUVX objects' variability parameters are plotted in the bottom row of
Figure~\ref{fig:AGs}. Estimating the completeness and purity in the S82
sampling  case, assuming that the 973 objects with spectra are a random
subset of all the nUVX objects, gives that we are 90\% complete and
96\% pure (Table~\ref{tab:compur}).  By the same argument as above, the
purity is an upper limit on the overall purity of the nUVX quasar
sample whereas the completeness is exact. The purity and the
completeness stand on their own in quantifying our ability to recover
the {\it spectroscopically confirmed quasars.}
Thus, the addition of the variability information enhances the purity
to 96\% instead of 8\%, as is the case for the pure color selected
sample. In the case of the sparser PS1-like sampling we still have a
completeness of 65\%, and a purity of 32\%. This clearly demonstrates
that a variability-based approach  is very efficient at selecting nUVX
quasars when the data is well sampled in time. Even with the sparse
\psone sampling, the purity increases by a factor of four when
variability information is used.

The plot of the S82-sampled nUVX objects (lower left corner of
Figure~\ref{fig:AGs}) shows a clear bimodality of the $A$ parameter
distribution of the unknown (red) objects. As seen with the object
training set, this bimodality is a probable separation between the
non-varying contaminants and the varying (possible RR~Lyrae) contaminants. Thus
the nUVX objects have been separated into quasar candidates (high
$\gamma$; intermediate $A$), RR~Lyrae candidates (low $\gamma$; high
$A$) and non-varying stars (low $\gamma$; low $A$).


\subsubsection{Quasars in the $griz$ Color Box}
\label{sec:grizboxres}

\begin{figure*}
\epsscale{0.70} 
\plotone{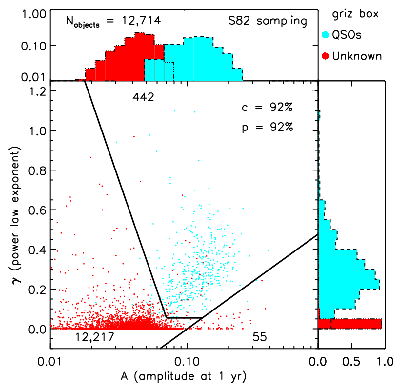}
\epsscale{0.99} 
\plottwo{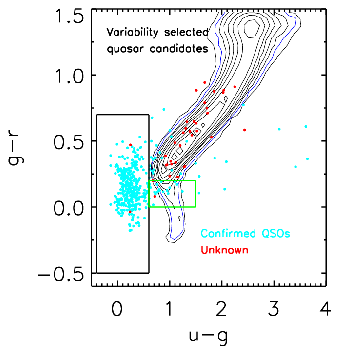}{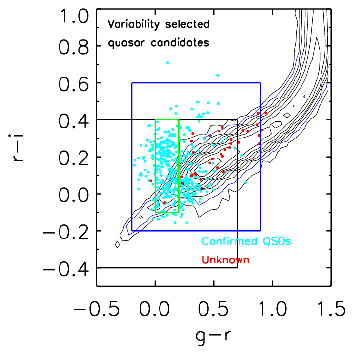}
\caption{The top panel shows the $A$ and $\gamma$ power law parameter
space of the objects in the $griz$ selection box
(Sections~\ref{sec:grizbox} and~\ref{sec:grizboxres}) with Stripe 82
time sampling. The projected probability distributions for $A$ and
$\gamma$ for the confirmed quasars (light blue points) and the unknown
objects (red points) are shown as histograms. The solid lines indicate
the selection cut defined in
Equations~\ref{eqn:cutDSone}--\ref{eqn:cutDSthree}. The numbers 12,217,
442 and 55 in the $A$-$\gamma$ plane indicate the number of objects in
the given region. The estimated completeness and purity is shown in the
upper right corner of the scatter plot. The bottom row shows the 442
variability selected quasar candidates (407 confirmed quasars and 35
unknown candidates) from the $A$-$\gamma$ space, projected back into
$ugr$ and $gri$ color space. The black, green and blue boxes correspond
to the UVX, nUVX and $griz$ selection boxes
(Sections~\ref{sec:UVXdata}--\ref{sec:grizbox} and
\ref{sec:UVX}--\ref{sec:grizboxres}). 87\% of the confirmed quasars and
6\% of the unknown candidates fall in the black UVX box in $ugr$ color
space (lower left plot). 5\% and 6\% of the quasars and unknowns fall
in the green nUVX box. The contours in the bottom plots indicate the
Stripe 82 stellar locus. 95\% of the stellar locus objects are within
the blue contour level (the last but one outer contour). Above 80\% of
the 35 unknown candidates and 41\% of the 407 confirmed quasars fall
within the blue $gri$ contour, providing an estimate of the number
of quasars ``hiding'' in the $gri$ stellar locus.}
\label{fig:AGcolor}
\end{figure*}

\begin{figure*}
\epsscale{0.70} 
\plotone{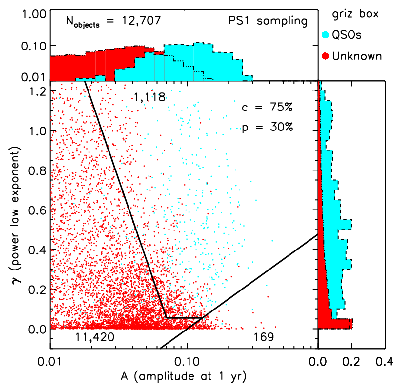}
\epsscale{0.99} 
\plottwo{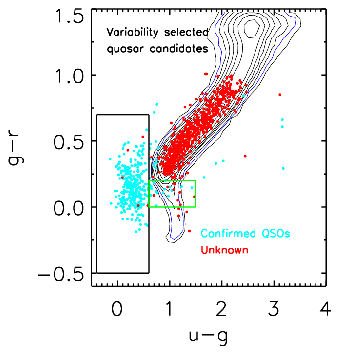}{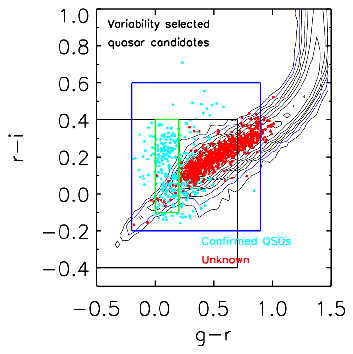}
\caption{Plots similar to those of Figure~\ref{fig:AGcolor} for the
$griz$ selection box data down-sampled to a PS1 cadence (6 epochs over
3 years). The numbers 11,420, 1,118 and 169 in $A$-$\gamma$ space (top
panel) indicate the number of objects in each of the three regions
defined by the black solid lines as defined in
Equations~\ref{eqn:cutDSone}--\ref{eqn:cutDSthree}. The estimated
completeness and purity of the 1,118 variability-selected quasar
candidates is shown in the upper right corner of the top panel. Of
these 1,118 objects, 42\% of the confirmed quasars and 98\% of the
unknown candidates fall within the blue $gri$ color space (lower right
plot) stellar locus contour level, which encloses 95\% of the Stripe 82
stellar locus objects. 87\% of the confirmed quasars fall in the black
UVX selection box in $ugr$ color space (lower left plot), as opposed to
only 1\% of the unknowns. 
In the green $ugr$ nUVX box fall 5\% and 1\% of the
quasars and unknown candidates. These percentages indicate that
most of the unknowns are stars scattered into our selection region due
to the 6 epoch sampling of PS1.}
\label{fig:AGcolorPS1}
\end{figure*}

To simulate quasar candidate selection without $u$-band photometry,  we
applied our variability analysis to the objects lying in a fairly
large multi-color region in  $griz$ space, the so-called $griz$ box
defined in Section~\ref{sec:grizbox}.
This color box fully contains an important part of the stellar locus.
Estimating the completeness and purity of our algorithm for the $griz$
box is difficult, as no clear estimates of the abundance of quasars
exist for such a color cut.  We therefore checked the quasar candidates
against a catalog of the SDSS+2SLAQ quasars, plus the 6 extra quasars
found via the visual inspection of the nUVX spectra. 
This may fall considerably short of a complete sample of quasars, but
at the moment it is the best we can do; the purities calculated in this
section are therefore lower limits.  Matching this quasar reference
catalog to the 12,714 objects in the $griz$ box we found 443 known
quasars.  Estimating $A$-$\gamma$ for all these sources, returned 442
(1,118)  variability-based quasar candidates, when considering the S82
(PS1) sampling and when applying the $A$-$\gamma$ cuts of
Equations~\ref{eqn:cutDSone}--\ref{eqn:cutDSthree}. Of these candidates
407 (333) were found to be known quasars. Thus, for the broad $griz$
color pre-selection, variability selection achieves 92\%(75\%) 
completeness and a purity of 92\%(30\%) for the S82 (PS1) sampled data,
respectively.  The $A$ and $\gamma$ distributions for the $griz$ box
selected objects are shown in the top panel of
Figures~\ref{fig:AGcolor} (S82 sampling) and  \ref{fig:AGcolorPS1} (PS1
sampling).

In the bottom panel of Figure~\ref{fig:AGcolor} and 
\ref{fig:AGcolorPS1}  we have projected our variability selected quasar
candidates back into $ugr$ and $gri$ color space, in order to
understand the color distributions of variability selected quasar
candidates. The figures show that 86\% and 98\%  of the not
spectroscopically confirmed candidates fall on the $gri$ stellar locus,
defined as  the (blue) contour level containing 95\% of the stellar
locus objects in S82. This suggests that many, if not most, of these
unconfirmed quasar candidates are  stars scattered into our
variability-selection region. However, there are some possible quasars
among the unknowns judging from their colors. For instance, a few of
the unknown objects (6\% and 1\% in the S82 and PS1 sampled case
respectively)  fall in the $ugr$ UVX selection box.  This illustrates
that our purity estimates are lower limits but close to the likely
truth.  
It also shows that the $ugriz$ quasar color
selection in SDSS \citep{richards02,richards06,richards09} has done an
excellent job, implying that $<10\%$ of
the quasars with $i<19.1$ are ``hiding''  in the stellar locus and have
been missed by the SDSS selection.

When defining the $griz$ box in Section~\ref{sec:grizbox} we included
the stellar locus to achieve as high a completeness as possible and to
illustrate the prospects of our approach. However, removing objects
falling within the stellar locus,  could greatly enhance the purity at
a modest reduction of the completeness.  Thus, selecting quasar
candidates without $u$-band information can be put into four scenarios:
\begin{itemize}

\item[1)] A ``naive'' $griz$ box ($i<19.1$) including the stellar locus
       and not considering variability information will have a quasar
       selection completeness above 90\%, but a purity of only 4\%. 

\item[2)] Taking a $griz$ box color selection, but cutting out the
      stellar locus (defined by the (blue) contour in
      Figures~\ref{fig:AGcolor} and \ref{fig:AGcolorPS1}) and still not
      considering variability information improves the purity to
      $\sim$48\% but lowers the completeness to $\sim$59\%. This is
      what would be easily achievable in a single-epoch $griz$ survey.

\item[3)] Combining a $griz$ color selection including the stellar
      locus, but considering variability information  (illustrated in
      Figures~\ref{fig:AGcolor} and \ref{fig:AGcolorPS1}) leads to a
      completeness of 92\%(75\%)  and a purity of 92\%(30\%) for the
      S82 (PS1) sampled data, respectively. This means that PS1 can 
      provide a 75\% complete quasar sample, if 30\% purity were
      acceptable.

\item[4)] Finally, removing the stellar locus from the $griz$ selection
      box and combining this with variability information lowers the
      completeness to 54\%(44\%), but returns a purity of 97\%(92\%)
      with S82 (PS1) time-sampled data. PS1 can provide a high purity
      quasar sample that is $\sim$50\% complete.

\end{itemize}
 
This illustrates how a variability selection cut greatly improves the
candidate sample as compared to quasar candidate selection based on
colors alone when $u$-band information is not available.  
Keeping in mind that the purities calculated here are lower limits, 
these results suggests that the variability selection of
quasar candidates in \psone or other multi-epoch surveys will produce
high quality, extensive samples.


\section{Discussion}
\label{sec:disc}

In this paper we have explored the selection of quasars by their
variability in lieu of their UV excess in the context of the PS1 $3\pi$
survey as a test case for various upcoming multi-epoch surveys.  Given
its extensive, multi-year time sampling, SDSS Stripe 82 is an
excellent test bed, that allows us to address variability selection of
quasars (and other sources) in general.
Besides \psone and SDSS, LSST \citep{ivezic08,LSST} and Gaia \citep{jordi06} 
will be able to employ variability selection
similar to that which we have presented. 

For any source, we have characterized its variability by a simple power
law model for its light curve structure function, which is similar to
(but not the same as) \cite{macleod08,sumi05,eyer02,cristiani96}. For
each object the amplitude~$A$ is the typical variability within 1 year,
and the exponent~$\gamma$ describes how the expectation value for the
magnitude differences changes with the time between measurements.
Applying this simple variability characterization to spectroscopically
confirmed quasars, to presumably non-variable sources (F/G stars) and
to known RR Lyrae, we have found that the $A$-$\gamma$ space separates
these source classes nicely. Quasars have $0.07<A<0.25$ and
$0.15<\gamma<0.5$; RR Lyrae have $A\sim0.2$ and $\gamma\sim0$, as the
rapid periodic variations in their light curves appear as ``white
noise'' with no secular trend when coarsely sampled on year-long time
scales; finally, non-variable sources have $A<0.05$. The variability
properties for quasars derived here for individual objects with
$\sim$60-epoch light curves, are consistent with those derived by for
instance \cite{vandenberk04} and \cite{bauer09} using ensemble
averaging. The results allowed the definition of a simple
variability-based quasar selection  in multi-epoch data, using cuts in
the $A$-$\gamma$ plane. As we summarize quantitatively below, this
variability selection works very well when considering 60-epoch light
curves in S82, and still works quite well for the sparser sampling
expected for PS1. 

The presented algorithm for variability selection is simple enough that
it can be applied to large samples. Our IDL code (not optimized for
speed) running on two dual-core CPUs takes 2 hours to characterize
the $\sim$12,000 objects of the $griz$ color box with the \psone
sampling.

The extensive spectroscopy of quasars in S82 has enabled stringent
completeness and purity estimates for variability selection, at least
for UVX quasars. Our analysis has shown that variability-based quasar
selection with only 6 epochs over 3 years (as for PS1) is effective,
producing either complete or pure samples. However, something more like 
the $\sim$60
epochs of S82 is needed to 
produce variability-selected quasar samples that are
both complete and pure (with only weak color pre-selection).

Therefore, it is paramount to eventually combine the variability
information in different  filters, in order to boost the number of
epochs in PS1. We can attempt to do this by predicting time-offset 
synthetic $r$-band magnitudes  from the $g$, $i$, $z$ and $Y$ bands
which could reduce the scatter in the $A$-$\gamma$  plane, and so
enhance the completeness and the purity of the \psone sampled
catalogs.  We leave this important extension of our method to further
work.

Considerably further in the future, LSST \citep{ivezic08,LSST} will start operating
and is planned to ultimately have $\sim$200 epochs spread over 10 years. 
This makes LSST optimal for creating
complete and pure variability selected quasar samples. Considering a UVX
box like the one used here, a quasar candidate variability selection with the LSST
cadence and a 10 year baseline would definitely push both the completeness
and purity above the (lower) S82 limits of 90\% and 95\% shown in
the top left plot of Figure~\ref{fig:AGs} and in Table~\ref{tab:compur}. 
Such samples are expected to be obtainable at least down to LSST's 10$\sigma$ 
limiting $i_\textrm{AB}$ magnitude of around 23.3 \cite{oguri10}, or even down to 5$\sigma$ 
($i_\textrm{AB}\lesssim24$) with the 200 epoch variability selection. 
If we further consider the fact that 
LSST will have a $u$-band, a LSST color-variability quasar selection will presumably 
be able to create (UVX) quasar catalogs almost as pure and complete as spectroscopic 
samples. Hence, LSST will be superior to PS1 in the 20,000 deg$^2$ LSST will 
observe. However, until LSST happens, PS1 is excellent for improving and further 
developing variability selection of quasars, and it will provide the first large variability
(and $grizy$-color) selected quasar samples.
The LSST (and PS1) quasar catalogs will with estimated sizes of 100 quasars per
square degree, or even more, continue the wild growth of 
quasar catalog sizes seen the last 40-50 years \citep{richards09}, and thereby stress 
that we are on the verge of an era where confirming the quasar nature of all objects 
in the catalogs by spectroscopic follow-up will be unfeasible. 
This makes the purity of the photometric samples crucial for
doing statistical analyses of the quasars. As demonstrated variability provides
purities above 90\% for well sampled data at all redshifts and not only for UVX
object and will therefore be an important step in achieving very pure photometric
quasar catalogs in the future. 


\section{Conclusions}
\label{sec:conc}

We have presented a simple, parameterized variability characterization
for sources with many epoch photometry. We do this by fitting a
power-law model for the structure function of each source's light
curve; the model is specified by an amplitude A and a power-law index
$\gamma$.

We have applied this approach to understanding variability-selection of
quasars in multi-epoch, multi-color surveys, either augmenting or
supplanting the more common color-selection. Specifically, we have 
analyzed data in SDSS Stripe 82 (S82) both to understand variability
selection {\it per se} and as a testbed for  ongoing and upcoming
surveys, such as PS1 and LSST. To predict PS1's ability to identify
quasars, we have down-sampled S82 data to a set of 6 epochs, resembling
the expected single-band time sampling in the PS1 3$\pi$ survey. 

For all sources in various sub-samples we calculated the parameters~$A$
and $\gamma$, and found that for sufficiently many epochs over a
multi-year interval, quasars separate  well in the $A$-$\gamma$
parameter plane from non-variable sources and e.g. RR Lyrae. Quasar
variability as typically characterized by $0.07<A<0.25$ and
$0.15<\gamma<0.5$, consistent with earlier ensemble analyses.

Drawing on the nearly complete spectral identification of quasars with
$i<19.1$ in S82, we have explored both the completeness and the purity
of single-band, variability-selected quasar samples with both S82 and
PS1 time-sampling,  and with or without the benefit of $u$-band
photometry (which PS1 will not have).

Specifically, we found the following:
\begin{itemize}

\item Among the complete, spectroscopically confirmed sample of
UV-excess quasars in S82, we can identify  90\% of them on the basis of
$\sim$60 $r$-band epochs over $\sim$5 years. This variability selected
sample only has a 5\% contamination from other objects.

\item Repeating the same exercise but with the 10-times sparser PS1
sampling, reduces the completeness to 73\% but with still high purity
(top right of Figure~\ref{fig:AGs}).

\item In the redshift range $2.5<z<3$, where quasars overlap with the
stellar locus in color-color space, variability can find 90\% (65\%) of
all spectroscopically confirmed quasars for S82 (PS1) sampling. This is
a factor of 5-10 more complete than existing color-selection in this
particular regime (bottom panel of Figure~\ref{fig:AGs}).

\item To understand our ability to select quasars through their
variability when no $u$-band data are available, we selected all
sources in a broad {\it griz} color box, known to contain almost all
spectroscopically confirmed quasars. In this color box, stars and other
contaminants outnumber quasars by a factor of 30 for $17<i<19.1$. 
Nonetheless, variability selection encompasses 92\% (75\%) of known
quasars for S82 (PS1) sampling. For S82 sampling
(Figure~\ref{fig:AGcolor}) the purity of this sample is still very high
(92\%), but it is rather lower with PS1-like sampling
(Figure~\ref{fig:AGcolorPS1}), 30\%. 

\item If in the case of PS1 (6 epochs, no $u$-band) a more pure quasar
sample is desired, this can be done by omitting the stellar locus in
the $griz$ box and then looking at the variability of the remaining
sources. This yields a completeness of only 44\%, but with a purity of
92\%.  

\item The high purity of the variability selected quasar sample in the
{\it griz} box (with S82 sampling) confirms that the fraction of
overlooked quasars in S82 must be small ($<10\%$); this inference is
predicted on the assumption that  the quasars ``buried'' in the stellar
locus have a  variability behavior similar to the others.

\item The same ($A$,$\gamma$) analysis is also very effective at
identifying RR Lyrae. With S82 sampling 97\% of the  RR Lyrae from
\cite{sesar09} are found, with PS1 sampling still 83\%; they appear as
objects with $(A,\gamma)\approx (0.2,0)$ (red points in
Figure~\ref{fig:AvsGallNOBIN} and \ref{fig:AvsGdsNOBIN}). 

\end{itemize}

As mentioned in the introduction, new and larger quasar samples have
many and  varied interesting  applications. One of these is the
possibility of finding lensed quasars. In this work we have only
focused on point source selection, with the purpose of finding quasar
candidates. Such an approach would be directly applicable to a search
for wide separation lenses, where (by definition) the multiple images
are well-resolved.   Another very interesting use of variability
selection, again with quasar lens finding in mind, is to look at
spatially extended objects rather than point sources. Applying the
algorithm to a set of extended, yet quasar-colored, objects will return
a list of small separation quasar lens candidates. We have initiated
such a search in Stripe 82 with promising preliminary results, which we
will present in a forthcoming paper.   This search is a pilot for the
\psone lensed quasar search, which will be able to be started after
about a year of $3\pi$ survey imaging has been built up. We expect to
find as many as 2000 lenses in this database \citep{oguri10}. Our
analysis illustrates that combining color-selection  of quasars with
variability selection is a powerful approach to take.


\section*{Acknowledgments}

We thank Gordon Richards, Xiao-hui Fan and David Hogg for useful discussions,
and Branimir Sesar for assistance with the RR~Lyrae sample.
KBS is funded by and would like to thank the Marie Curie Initial Training
Network ELIXIR, which is funded by the Seventh Framework Programme (FP7) of
the European Commission.
PJM was supported in part by research fellowships from the TABASGO and Kavli
foundations, and by the US Department of Energy under contract number
DE-AC02-76SF00515.

Funding for the SDSS and SDSS-II has been provided by the Alfred P. Sloan
Foundation, the Participating Institutions, the National Science Foundation,
the U.S. Department of Energy, the National Aeronautics and Space
Administration, the Japanese Monbukagakusho, the Max Planck Society, and the
Higher Education Funding Council for England. The SDSS Web Site is
\verb+http://www.sdss.org/+.

The SDSS is managed by the Astrophysical Research Consortium for the
Participating Institutions. The Participating Institutions are the American
Museum of Natural History, Astrophysical Institute Potsdam, University of
Basel, University of Cambridge, Case Western Reserve University, University of
Chicago, Drexel University, Fermilab, the Institute for Advanced Study, the
Japan Participation Group, Johns Hopkins University, the Joint Institute for
Nuclear Astrophysics, the Kavli Institute for Particle Astrophysics and
Cosmology, the Korean Scientist Group, the Chinese Academy of Sciences
(LAMOST), Los Alamos National Laboratory, the Max-Planck-Institute for
Astronomy (MPIA), the Max-Planck-Institute for Astrophysics (MPA), New Mexico
State University, Ohio State University, University of Pittsburgh, University
of Portsmouth, Princeton University, the United States Naval Observatory, and
the University of Washington.


\begin{appendix}

\section{Individual structure function parameter inference by MCMC}
\label{sec:mcmc1}

We use a simple Markov chain Monte Carlo (MCMC) approach 
\citep[e.g.][]{metropolis53,hastings70,press92,hansen03} to  infer the
parameters $A$ and $\gamma$ (Equation~\ref{eqn:powerlaw}) and their
confidence regions. Our MCMC procedure takes as input a catalog of 
magnitudes, photometric errors and observed frame MJDs, converted to
$N(N-1)/2$ pairs of observations (Equation~\ref{eqn:datapairs}).  We
then initialize it as follows:
\begin{itemize}
 \item Pick a starting point for $A$ and $\gamma$: we chose 0.1 for both
 \item Define an initial Gaussian proposal distribution (PD)
 \item Set an initial temperature~$\beta$ for the chain
\end{itemize}
The PD width sets the mobility of the Markov chain. Based on several
tests we set the initial PD width to 0.05 in both the $\log{A}$ and
$\gamma$ directions.  During a ``burn-in'' period at the start of
sampling, we draw samples from a modified posterior PDF for the 
parameters, given by the
product of the prior PDF, and the likelihood raised to the power
of~$\lambda$. This parameter is an inverse temperature; we start from
$\lambda_0=\frac{1}{\beta_0}=10^{-3}$.  The $\lambda$ parameter is then
increased geometrically to unity as 500 samples are drawn, at which
point burn-in is declared over, the inverse temperature is fixed at
$\lambda=1$, and the subsequent samples are stored and used to compute
various statistics. For the post burn-in sampling, we use an updated
Gaussian proposal distribution, whose widths are set to 10\% of the 
standard deviations  of the parameter ($\log{A}$  and~$\gamma$) values
sampled during burn-in.

At each point in parameter space proposed, we calculate the
(un-normalized) log posterior probability distribution of the step,
which we define as
\begin{equation}\label{eqn:logP}
	\log P = \log P(A) + \log P(\gamma) - \lambda \frac{\chi^2}{2} \; .
\end{equation}
Here, $\chi^2$ is related to the logarithm of the likelihood defined in
Equation~\ref{eqn:L}
\begin{equation}\label{eqn:logL}
  \chi^2 = -2 \log \mathcal{L} =  
     \left(   \sum_{ij} \log( 2\pi V_{{\rm eff},ij}^2) + 
        \sum_{ij} \frac{\Delta m_{ij}^2}{V_{{\rm eff},ij}^2} \right) \; ,
\end{equation}
with the effective variability defined as in Equation~\ref{eqn:Veff}:
\begin{equation}
V_{{\rm eff},ij}^2 =
  V_{\rm mod}(A,\gamma | \Delta t_{ij})^2 + \delta \Delta m_{ij}^2 
  = \left(A \Delta t_{ij}^{\gamma}\right)^2 + \delta \Delta m_{ij}^2 \; .
\end{equation}
The sums in Equation~\ref{eqn:logL} are over the data pairs derived 
from the light curve of the given object, and $\delta\Delta m_{i,j}$
is the photometric error on the $ij^{\rm th}$~magnitude pair.
In Equation~\ref{eqn:logP} $\log P(A)$ and $\log P(\gamma)$  represent the
(log) prior PDFs for the parameters $A$ and $\gamma$, which we chose to be uninformative.
We assigned the following functional forms:
\begin{eqnarray}
P(A) &\propto& \frac{1}{A} \\
P(\gamma) &\propto& \frac{1}{1+\gamma^2} \; .
\end{eqnarray}
If $\gamma$ is negative or $A$ lies outside the range $[0,1]$,  the log
prior density for that parameter is set to $-10^{32}$,  lowering the
overall posterior probability for that particular iteration step to
effectively zero.  In this way we enforce our assumption that the power
law exponent is positive and that the average variability on a 1 year
timescale is less than 1 magnitude \citep[as found by e.g.][]
{vandenberk04,bauer09}. Samples are accepted or rejected via the
Metropolis-Hastings algorithm \citep{metropolis53,hastings70}. Care is
taken to make the comparison of the log posterior values at constant
inverse temperature.

For our final $A$ and $\gamma$ values, we choose to take the position
of the global peak of the likelihood, an approximation of the
``best-fit'' point.  We approximate this by keeping track of the value
of the likelihood  as we sample, and then using the sample with the
highest value as our estimate. In practice the posterior PDF is not
dominated by the prior,  such that the peaks of the likelihood and the
posterior PDF are usually quite close together.

The uncertainties on the parameters are estimated  by considering the
$16^{\rm th}$ and $84^{\rm th}$ percentiles of the 1-dimensional
marginalized distributions. In the case of a (symmetric) Gaussian
distribution, this would correspond to the $1\sigma$ error bar.

To confirm our choice of initial conditions, cooling schedule and PD
evolution as sensible, we tested our sampler on simulated light curves
for both sine wave and Gaussian white noise sources,  and recovered the
correct parameters (zero $\gamma$ and analytically calculated~$A$).

\end{appendix}

\end{document}